# Optimal Power Flow Solutions via Noise-Resilient Quantum-Inspired Interior-Point Methods

Farshad Amani, *Student Member, IEEE*, Amin Kargarian, *Senior Member, IEEE*

*Abstract*—This paper presents three quantum interior-point methods (QIPMs) tailored to tackle the DC optimal power flow (DCOPF) problem using noisy intermediate-scale quantum devices. The optimization model is redefined as a linearly constrained quadratic optimization. By incorporating the Harrow-Hassidim-Lloyd (HHL) quantum algorithm into the IPM framework, Newton's direction is determined through the resolution of linear equation systems. To mitigate the impact of HHL error and quantum noise on Newton's direction calculation, we present a noise-tolerant quantum IPM (NT-QIPM) approach. This approach provides high-quality OPF solutions even in scenarios where inexact solutions to the linear equation systems result in approximated Newton's directions. Moreover, to enhance performance in cases of slow convergence and uphold the feasibility of OPF outcomes upon convergence, we propose a hybrid strategy, classically augmented NT-QIPM. This technique is designed to expedite convergence relative to classical IPM while maintaining the solution accuracy. The efficacy of the proposed quantum IPM variants is studied through comprehensive simulations and error analyses on 3-bus, 5-bus, 118-bus, and 300-bus systems, highlighting their potential and promise in addressing challenging OPF scenarios. By modeling the errors and incorporating quantum computer noise, we simulate the proposed algorithms on both Qiskit and classical computers to gain a deeper understanding of the effectiveness and feasibility of our methods under realistic conditions.

## I. INTRODUCTION

OPTIMAL power flow (OPF) is a fundamental problem in power systems that serves as a core for various energy management functions. OPF aims to ensure the continuous supply of electricity while optimizing the cost of power systems. It is central to several power systems operation problems such as demand response, unit commitment, reliability, stability, resilience, etc. OPF is solved several times every day. With the increasing complexity of power systems, classical optimization algorithms are becoming inadequate in terms of computational efficiency and speed. This has motivated exploring quantum computation integration into optimization algorithms, particularly in future power grids under emerging technologies [1-5]. Quantum computers are expected to provide a computational advantage that surpasses the capabilities of the most advanced classical computers.

OPF is an optimization problem that has both continuous and discrete variables characterized by nonlinearity, non-convexity, and large-scale complexity. Many OPF methodologies have been suggested based on linear programming [6], nonlinear programming [7], quadratic programming [8], heuristic methods [9], dynamic programming [10], lambda iteration method [11], and most recently machine learning methods [12]. The complexity of each of these methods is different. Although a full AC optimal power flow (ACOPF) formulation better represents a power system, it is non-convex and computationally complex [13]. DC optimal power flow (DCOPF) can provide a desirable solution that satisfies basic system information [14]. Various mathematical methods have been developed for solving DCOPF, including gradient-based methods, interior point methods (IPM), and sequential quadratic programming [1], [15]. In these methods, we face a system of linear equations that should be solved to obtain the OPF solution.

Most of these early approaches for solving complex OPF problems had slow convergence and limited applicability. [16] introduces significant advancement in OPF. Karush-Kuhn Tucker optimality conditions are solved directly using the Newton method and sparsity techniques to accelerate computations. However, this approach faced challenges in identifying active inequality constraints. Fiacco and McCormick addressed this problem by developing IPM [17]. IPM outperforms the classical simplex method, particularly for large linear programming problems [18]. Primal-dual and predictor-corrector IPM proposed in [19] serves as the benchmark.

IPM is categorized into two groups: feasible and infeasible, depending on the feasibility of the initial point [15]. At each iteration of IPM variants, a system of linear equations must be solved to find the search direction or Newton step. Traditionally, this system of linear equations is solved using Bunch-Parlett factorization in the case of symmetric indefinite or Cholesky factorization for positive definite matrices [20]. Multiple methods, such as Krylov subspace, are presented to solve the linear system of equations inexactly but with fewer iterations. While the fastest practical classical algorithm to solve a system of linear equations has a time complexity of $O(n^3)$, it is possible to solve it in a logarithmic time complexity by taking advantage of quantum computation abilities [1]. Harrow-Hassidim-Lloyd (HHL) is one of the quantum computation algorithms that can solve linear systems of equations at an exponential speedup compared to classical methods [22]. HHL finds all the matrix eigenvalues using the phase estimation method and then rotates them to find the solution. This process is done by a few quantum gates that make

This work was supported by the National Science Foundation under Grants ECCS-1944752 and ECCS-2312086.

The authors are with the Department of Electrical and Computer Engineering, Louisiana State University, Baton Rouge, LA 70803 USA (e-mail: famani1@lsu.edu, kargarian@lsu.edu).

HHL very fast [23].

HHL has gained extensive attention from researchers in different areas of study due to its salient feature in solving linear systems of equations. HHL has been used in power systems to perform Newton-Raphson iterations for solving AC and DC power flow problems [24], [25]. These two references lay the foundation for implementing HHL in addressing fundamental calculations in power system analysis. [26] explores the potential of HHL for solving DC power flow through a hybrid quantum-classical approach and evaluates its error and accuracy compared to the conventional HHL method. The authors of [27] utilize HHL to overcome the challenges associated with solving discrete-time nodal equations in electromagnetic transients, which is a complex problem for conventional simulators. In [28], the authors address a unit commitment problem using a combination of HHL and quantum approximate optimization algorithms. This hybrid approach demonstrates a cubic speedup compared to classical algorithms.

Even though HHL exhibits fast computational speed, both quantum computers and HHL encounter errors due to their operation in the noisy intermediate-scale quantum. Hence, we expect an erroneous solution from HHL. Using HHL to calculate Newton's direction, we must ensure that IPM remains in a feasible region. [29] suggests a general inexact feasible IPM. Since HHL is an inexact system of linear equations solver, it is necessary to ensure that the feasibility of IPM will be maintained while using this solver. [30] provides a method that guarantees feasibility even under an inexact system of linear equations solver.

This paper introduces three Quantum IPM (QIPM) algorithms designed to address the DCOPF problem efficiently using noisy intermediate-scale quantum devices. The core optimization task is reframed as a linearly constrained quadratic optimization. Integrating the HHL quantum algorithm into the IPM framework enables the calculation of Newton's direction by solving linear equation systems. To counteract the influence of HHL error and quantum noise on Newton's direction calculation, we present a noise-tolerant Quantum IPM (NT-QIPM) strategy. This methodology guarantees high-quality OPF solutions even when faced with inexact solutions to the linear equation systems, potentially leading to approximated Newton's directions. To further enhance performance in cases of sluggish convergence and ensure the feasibility and exactness of OPF results upon convergence, we propose a hybrid technique named classically augmented NT-QIPM (CNT-QIPM). This method is developed to theoretically accelerate convergence compared to classical IPM while maintaining identical solution accuracy. We demonstrate the promising capabilities of the newly proposed quantum IPM variants through simulation outcomes and error analyses.

## II. CLASSICAL INTERIOR POINT METHOD DCOPF

The DCOPF problem is formulated in (1)-(7). The objective is the minimization of quadratic generation costs. Equality constraints are nodal power balance (2), line flow equations (3), and voltage angle of the reference bus (4). Inequality constraints include generation capacity limits (5), upper and lower bounds of voltage angles (6), and line flow limits (7).

$$\min \mathcal{C}^\dagger p_g + \frac{1}{2} p_g{}^\dagger \mathcal{Q} p_g \quad (1)$$

$$\sum_g p_{gi} - P_{di} = \sum_j p_{ij} \quad \forall i \quad (2)$$

$$p_{ij} = \frac{\theta_i - \theta_j}{x_{ij}} \quad \forall ij \quad (3)$$

$$\theta_{ref} = 0 \quad (4)$$

$$\underline{P}_g \le p_g \le \overline{P}_g \quad \forall g \quad (5)$$

$$-\pi \le \theta_i \le \pi \quad \forall i \quad (6)$$

$$-\overline{P}_{ij} \le p_{ij} \le \overline{P}_{ij} \quad \forall ij \quad (7)$$

where $p_g$ is power generation of unit $g$. $\mathcal{C}$ and $\mathcal{Q} \in \mathbb{R}^{n \times n}$ are cost function coefficients. $\mathcal{C}^\dagger$ denotes the transpose of $\mathcal{C}$. Matrix $\mathcal{Q}$ is a square matrix with non-zero elements only in its main diagonal; therefore, it is symmetric and positive semidefinite. $\underline{P}_g$ and $\overline{P}_g$ are the minimum and maximum generator capacity, $\theta_i$ represents the angle of voltage at bus $i$, $p_{ij}$ denotes power flow in line between buses $i$ and $j$, $\overline{P}_{ij}$ is the maximum allowed line flow, and $P_{di}$ is the amount of load at node $i$. Susceptance of line $ij$ is shown by $x_{ij}$.

Using slack variables, we turn inequalities into equality constraints:

$$p_g + \bar{s}_g = \overline{P}_g \quad \forall g \quad (8)$$

$$-p_g + \underline{s}_g = \underline{P}_g \quad \forall g \quad (9)$$

$$\theta_i + \bar{s}_\theta = \pi \quad \forall i \quad (10)$$

$$-\theta_i + \underline{s}_\theta = \pi \quad \forall i \quad (11)$$

$$p_{ij} + \bar{s}_L = \overline{P}_{ij} \quad \forall ij \quad (12)$$

$$-p_{ij} + \underline{s}_L = \overline{P}_{ij} \quad \forall ij \quad (13)$$

$$\bar{s}_g, \underline{s}_g, \bar{s}_\theta, \underline{s}_\theta, \bar{s}_L, \underline{s}_L \ge 0 \quad \forall g, i, ij \quad (14)$$

This transformation leads to the minimization of the problem, as outlined in (1), while adhering to the conditions specified in (2)-(4) and (8)-(14). The equality constraints are represented in a compact form as $\mathcal{G} x^\mathcal{P} = \mathcal{I}$, in which:

$$\mathcal{G} x^\mathcal{P} = \mathcal{I} \Rightarrow$$

$$\begin{pmatrix} \mathcal{S}_g & 0 & 0 & \mathcal{S}_s^g \\ 0 & \mathcal{S}_\theta & 0 & \mathcal{S}_s^\theta \\ 0 & 0 & \mathcal{S}_{p_{ij}} & \mathcal{S}_s^{p_{ij}} \end{pmatrix} \begin{pmatrix} P_g^\mathcal{P} \\ \theta_i \\ p_{ij} \\ s \end{pmatrix} = \begin{pmatrix} (\overline{P}_g, -\underline{P}_g)^T \\ (\pi, -\pi)^T \\ (\overline{P}_{ij}, -\overline{P}_{ij})^T \end{pmatrix} \quad (15)$$

$\mathcal{S}_s^g \in \mathbb{R}^{2g \times g}$ represents slack variables corresponding to generation limits in (8) and (9). $\mathcal{S}_\theta \in \mathbb{R}^{2(b-1) \times (b-1)}$ shows (10) and (11) and $\mathcal{S}_s^\theta \in \mathbb{R}^{2b \times b}$ are their slack variables. $\mathcal{S}_{p_{ij}} \in \mathbb{R}^{2L \times L}$ is for line constraints with slack variables $\mathcal{S}_s^{p_{ij}} \in \mathbb{R}^{2L \times L}$.

Now, we represent the DCOPF formulation as linearly constrained quadratic optimization that involves optimizing the quadratic objective function (1) subject to inequality (14) and linear constraints (15). The general primal form of DCOPF ($\mathcal{P}$) and the dual linearly constrained quadratic DCOPF ($\mathcal{D}$) are represented as:

$$(\mathcal{P}): \quad \min \mathcal{C}^\dagger x^\mathcal{P} + \frac{1}{2} x^{\mathcal{P}\dagger} \mathcal{Q} x^\mathcal{P} \quad (16)$$

$$\text{s.t. } \mathcal{G} x^\mathcal{P} = \mathcal{I}$$

$$(\mathcal{D}): \quad \begin{array}{c} x^{\mathcal{P}} \geq 0 \\ \max \mathcal{J}^{\dagger} y^{\mathcal{D}} - \frac{1}{2} x^{\mathcal{P}\dagger} Q x^{\mathcal{P}} \\ \text{s.t. } \mathcal{G}^{\dagger} y^{\mathcal{D}} + s^{\mathcal{D}} - Q x^{\mathcal{P}} = \mathcal{C} \\ s^{\mathcal{D}} \geq 0 \end{array} \quad (17)$$

In $\mathcal{P}$, we aim to find the values of vector $x^{\mathcal{P}} = (p_g, \theta_i, p_{ij}, s)^T$ that minimize generation cost (1), given that linear constraints defined by matrix $\mathcal{G} \in \mathbb{R}^{m \times n}$ and vector $\mathcal{J} \in \mathbb{R}^m$ are satisfied. $m$ is the number of equality constraints, and $n$ is the number of variables in $x^{\mathcal{P}}$. In $\mathcal{D}$, $y^{\mathcal{D}} \in \mathbb{R}^m$ and $s^{\mathcal{D}} \in \mathbb{R}^n$ correspond to dual variables associated with the primal problem. If the set of interior feasible solutions can be expressed as $\mathcal{PD}$ set, based on the principle of strong duality, the optimal solution can be characterized by $\mathcal{PD}^* := \{(x^{\mathcal{P}}, y^{\mathcal{D}}, s^{\mathcal{D}}) \in \mathcal{PD} : x^{\mathcal{P}} \circ s^{\mathcal{D}} = 0\}$. Assuming $\epsilon > 0$, the set of $\epsilon$-approximate solutions for problem (1) is expressed as:

$$\mathcal{PD}_\varepsilon := \left\{ (x^{\mathcal{P}}, y^{\mathcal{D}}, s^{\mathcal{D}}) \in \mathcal{PD} : x^{\mathcal{P}\dagger} s^{\mathcal{D}} \leq n\varepsilon \right\}. \quad (18)$$

Consider diagonal matrices $\mathbb{X}$ and $\mathbb{S}$, where $\mathbb{X}$ represents the diagonal matrix of $x^{\mathcal{P}}$ and $\mathbb{S}$ represents the diagonal matrix of $s^{\mathcal{D}}$. For any positive value of $\mu$, the perturbed optimality conditions have a unique solution denoted by $(\mathcal{X}^{\mathcal{P}}(m), \mathcal{Y}^{\mathcal{D}}(m), \mathcal{S}^{\mathcal{D}}(m))$, which defines the central path for both the primal and dual problems:

$$\mathcal{CD} := \{(\mathcal{X}^{\mathcal{P}}, \mathcal{Y}^{\mathcal{D}}, \mathcal{S}^{\mathcal{D}}) \in \mathcal{PD}^0 | \mathcal{X}^{\mathcal{P}}_i \mathcal{S}^{\mathcal{D}}_i = \mu \text{ for } i \in \{1, \ldots, n\}; \text{ for } \mu > 0\}. \quad (19)$$

The central path ensures that the algorithm remains inside the feasible region throughout its iterations, allowing for efficient convergence to the optimal solution. Classical IPM utilizes Newton's method to solve perturbed optimality conditions for finding the Newton direction in each iteration of IPM. An updated candidate solution for the primal-dual linearly constrained quadratic optimization pair in (16) and (17) is obtained by solving the following linear system:

$$\begin{bmatrix} \mathcal{G} & 0 & 0 \\ -Q & \mathcal{G}^T & I \\ \mathbb{S} & 0 & \mathbb{X} \end{bmatrix} \begin{bmatrix} \Delta x^{\mathcal{P}} \\ \Delta y^{\mathcal{D}} \\ \Delta s^{\mathcal{D}} \end{bmatrix} = \begin{bmatrix} r_p \\ r_d \\ r_c \end{bmatrix} = \begin{bmatrix} \mathcal{J} - \mathcal{G} x^{\mathcal{P}} \\ \mathcal{C} - \mathcal{G}^{\dagger} y^{\mathcal{D}} - s^{\mathcal{D}} \\ \sigma \mu e - \mathbb{X}\mathbb{S}e \end{bmatrix} \quad (20)$$

where $\sigma \in (0,1)$ represents the extent of barrier reduction, and $(r_p, r_d, r_c)$ are the residuals. When $r_p = 0$ and $r_d = 0$, the solution $(x^{\mathcal{P}}, y^{\mathcal{D}}, s^{\mathcal{D}})$ is primal and dual feasible. Solving system-of-linear equations (20) is one of the most time-consuming calculations in IPM. We aim to tackle this challenge and reduce IPM running time by leveraging quantum computation techniques.

## III. QUANTUM INTERIOR POINT METHOD

Defining $\mathcal{A} = \begin{bmatrix} \mathcal{G} & 0 & 0 \\ -Q & \mathcal{G}^T & I \\ \mathbb{S} & 0 & \mathbb{X} \end{bmatrix}$ and residual vector $\mathcal{R} = \begin{bmatrix} r_p \\ r_d \\ r_c \end{bmatrix}$, we show (20) as $\mathcal{AX} = \mathcal{R}$. Instead of solving this system with common classical algorithms, we pass this linear system of equations to HHL, a quantum computing algorithm. HHL has a time complexity of $O(N s_0 \kappa \log(1/\varepsilon_0))$. This complexity analysis incorporates the matrix dimensions ($N$), sparsity ($s_0$), condition number ($\kappa$), and the desired error bound ($\varepsilon_0$). For solving (20) with HHL, a dedicated circuit is needed. The HHL circuit, shown in Fig. 1, consists of three registers of qubits. The ancilla register initialized at $|0\rangle_{Anc}$ state is responsible for eigenvalue rotations. The work registers initialized at $|0\rangle_\omega$ state hold eigenvalues of square matrix $\mathcal{A}$ obtained from the phase estimation process. Additionally, registers are used for encoding the classical vector $\mathcal{R}$ into a quantum state $|\tilde{r}\rangle$, which in this paper is the vector of IPM residuals. The number of work registers relies on the intended degree of precision for eigenvalues.

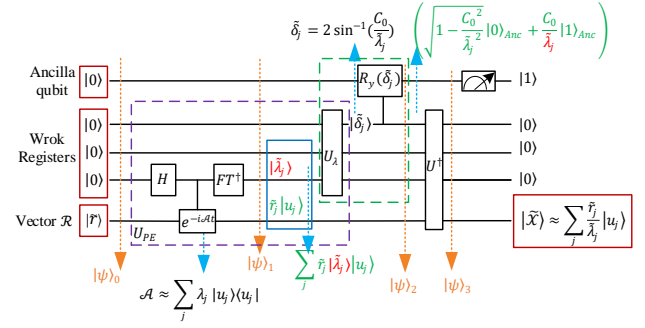

Fig. 1. HHL circuit [1].

The process of obtaining the solution of a system of linear equations by quantum computers is as follows. First, the classical vector $\mathcal{R}$ is encoded into a quantum state $|\tilde{r}\rangle$. This step prepares the quantum representation of the input vector. Then, the quantum phase estimation stage estimates the eigenvalues ($|\tilde{\lambda}_j\rangle$) of matrix $\mathcal{A}$. This involves applying controlled operations that effectively extract eigenvalues into the set of work registers qubits. At this stage, the entangled state $|\psi\rangle_1 = \sum_j |0\rangle_{Anc} \otimes \tilde{r}_j |\tilde{\lambda}_j\rangle_\omega \otimes |u_j\rangle$, in which $|u_j\rangle$ denotes the normalized eigenbases of matrix $\mathcal{A}$, is available. To invert the phase information of the work register, the HHL circuit performs rotations on eigenvalues. This stage involves applying specific quantum gates using a y-axis rotation on the ancilla qubit. At this point, we will obtain $|\psi\rangle_2$:

$$|\psi\rangle_2 = \sum_j \left( \sqrt{1 - \frac{C_0^2}{\tilde{\lambda}_j^2}} |0\rangle_{Anc} + \frac{C_0}{\tilde{\lambda}_j} |1\rangle_{Anc} \right) \otimes \tilde{r}_j |\tilde{\lambda}_j\rangle_\omega \otimes |u_j\rangle \quad (21)$$

The next step is performing inverse quantum Fourier transform on work registers to convert the phase information back into amplitudes, allowing for further calculations. By uncomputing work registers, this process removes redundant information and resets them to $|0\rangle$ state. $|\psi\rangle_3$ is the entangled state that we obtain at this point.

$$|\psi\rangle_3 = |0\rangle_\omega \otimes \sum_j \left( \sqrt{1 - \frac{C_0^2}{\tilde{\lambda}_j^2}} |0\rangle_{Anc} + \frac{C_0}{\tilde{\lambda}_j} |1\rangle_{Anc} \right) \otimes \tilde{r}_j |u_j\rangle \quad (22)$$

We can see that work registers are turned back to $|0\rangle_\omega$. For solution extraction, measurements are performed on the state $|1\rangle$ from ancilla qubit to obtain eigenvalues' amplitudes. These amplitudes contain the information needed to reconstruct the solution of the linear equations.

$$|\psi\rangle_{sol} = C_0 \sum_j \frac{\tilde{r}_j}{\tilde{\lambda}_j} |u_j\rangle \quad (23)$$

$|\psi\rangle_{sol}$ is the desired state for vector $|\tilde{x}\rangle = \sum_j \frac{\tilde{r}_j}{\tilde{\lambda}_j} |u_j\rangle$. State $|\tilde{x}\rangle$

can be converted into the classical $\mathcal{X}$ using a normalization proportionality factor. Vector $\mathcal{X}$ is the answer to the system of linear equations (20) and is considered an iteration of IPM. The general algorithm of HHL is presented in Algorithm I.

---

**Algorithm I** HHL.

**Input:** Matrix $\mathcal{A}$ and column vector $\mathcal{R}$
**Output:** A solution vector $\mathcal{X}$
1: Encode residual vector $\mathcal{R} = (r_p, r_d, r_c)^T$ as $|\tilde{r}\rangle$
2: Initialize unitary operation $\mathcal{U} = e^{-i\mathcal{A}t}$
3: Perform quantum phase estimation
4: Applied controlled rotation $R_y(\tilde{\delta}_j)$
5: Uncompute the work resisters using inverse quantum Fourier transform
6: Measure state $|1\rangle$ of ancilla qubit
8: Using normalization proportionality factor, convert $|\tilde{\mathcal{X}}\rangle$ to classical vector $x$
9: **Return** $x = (x^{\mathcal{P}}, y^{\mathcal{D}}, s^{\mathcal{D}})$

---

## III. IMPACTS OF QUANTUM NOISE AND ERROR

HHL is susceptible to a range of errors and inaccuracies, which can be classified into two categories: algorithm-related inaccuracies and quantum hardware-related errors. These errors can lead to undesired inaccuracies in the obtained results. This section aims to provide an understanding of the type and impact of these errors on DCOPF when solved by QIPM.

### A. Quantum Hardware Noise

Quantum hardware-related errors, such as quantum gates, lack of sufficient qubits connectivity, or measurement errors, are due to imperfections in the physical components of quantum computers. Due to noises and gate imperfection, the accuracy of HHL and QIPM is affected. This situation worsens by a lack of enough connectivity between qubits of the current quantum processor unit, which ended in the transpile of the circuit to a larger one; hence, more gates and more errors. These errors result in incorrect solutions vector $\mathcal{X}$.

Qiskit, an open-source software platform developed by IBM, provides a convenient feature that automatically generates a basic noise model for IBMQ hardware devices [31]. This noise model can be utilized to conduct simulations of quantum circuits, enabling us to implement the impact of quantum hardware errors on real quantum devices. However, these models are approximations of actual errors experienced on physical devices based on a limited set of input parameters with average error rates on gates.

For comparing the output of the real device with the noisy output, consider a 4-qubit state $\frac{1}{\sqrt{2}}(|0,0,0,0\rangle + |1,1,1,1\rangle)$. Before running with noise or on the device. The ideal expected output with no noise is as follows:

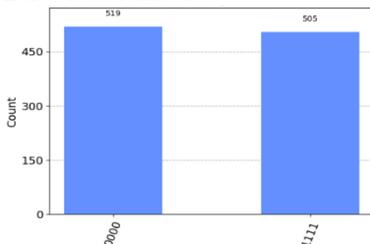

Fig. 2 Ideal counts for 4-qubits GHZ state.

The described noise model considers errors such as:
- Errors in single-qubit gates, which affect the accuracy of operations on individual qubits.
- Errors in two-qubit gates, which introduce noise and imperfections in two-qubit operations.
- Errors in readout, which impact the reliability of measurement results.

The presence of noise has a distinct impact, typically leading to errors and discrepancies in the resultant output. This phenomenon becomes evident when comparing Fig. 3 with the anticipated outcomes presented in Fig. 2. Specifically, Fig. 3 highlights the emergence of states beyond the desired ones, introducing errors into the output. These extra undesired states, originating from noise and system imperfections, significantly contribute to deviations from the ideal output.

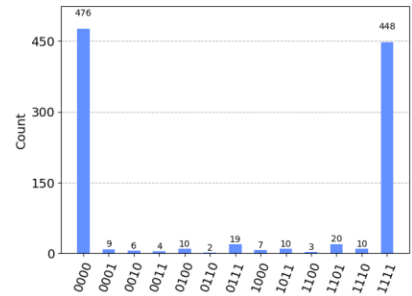

Fig. 3 Counts for 4-qubits GHZ state with noise model.

### B. HHL Algorithm Error

Algorithm-related inaccuracies refer to errors arising from inherent limitations in the HHL algorithm. These inaccuracies can stem from factors such as the choice of encoding classical data methods, limited work registers qubits, or estimating eigenvalues instead of their exact values. For instance, HHL calculates an approximate inverse of the input matrix $\mathcal{A}$ (in $\mathcal{AX} = \mathcal{R}$), which can introduce errors if the approximation is not accurate enough. In addition, the process of encoding matrix $\mathcal{A}$ as a unitary quantum gate $\mathcal{U} = e^{-i\mathcal{A}t}$ is an essential step of HHL, while it is prone to errors. These errors result in a deviation between the unitary gate $\mathcal{U}$ and the exact classical matrix $\mathcal{A}$.

Various methods can code matrix $\mathcal{A}$ as a unitary gate $\mathcal{U}$, each possessing a different precision. Hamiltonian simulation is one of these methods that can encode matrix $\mathcal{A}$ while $\|\mathcal{U} - e^{-i\mathcal{A}t}\| < \varepsilon_{\mathcal{A}}\|$. $\varepsilon_{\mathcal{A}}$ represents the maximum deviation of the unitary gate with the exact value.

## IV. NOISE MITIGATION

### A. Noise Tolerant Quantum IPM (NT-QIPM)

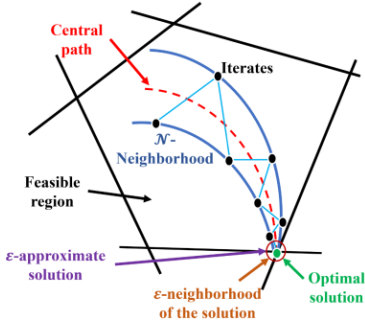

Fig. 4 IPM central path over iterations.

The process that a classical IPM takes to find the optimal solution is shown in Fig. 4. IPM identifies the Newton direction within a central path defined by (19) and progresses along that direction toward the optimal solution. However, this movement is restricted to a limited neighborhood around the central path. This neighborhood is determined by the value of step size $\mu$ in (19). This trajectory (blue curves in Fig. 4) lies within a region where both the primal and dual feasibility conditions are satisfied, and it gradually approaches the optimal solution while maintaining these feasibility conditions. Getting an inexact solution for the linear system (20) leads to inexact infeasible IPM. However, infeasible solutions may damage power system components if an operator makes any decision based on those values. For instance, when active power flows exceed the thermal capacity, it can damage transmission lines. Therefore, it is essential to ensure that QIPM provides feasible results despite quantum noise and error.

The central path neighborhood is as follows:
$$\mathcal{N}(\theta) \coloneqq \{(x^{\mathcal{P}}, y^{\mathcal{D}}, s^{\mathcal{D}}) \in \mathcal{PD}^0 | \ \|\mathbb{X}\mathbb{S}e - \mu e\| \leq \theta\mu\}, \quad (24)$$
where $\theta \in (0,1)$. If step size $\mu$ remains within the feasible region or a neighborhood surrounding the central path, the solution would be feasible.

Due to the inability of HHL to provide exact solutions in each QIPM iteration, the final solution can be erroneous. We propose a noise-tolerant QIPM (NT-QIPM) that guarantees to remain within the feasible region and attain a desirable solution even under the inexact solution of the system of linear equations. It is possible to reformulate (20) in a way that the value of $r_p$ and $r_d$ become zero and the effect of inexactness appears in the third residual value $r_c$ as follows [30]:
$$\mathbb{S}V\lambda + \mathbb{X}(QV\lambda - \mathcal{G}^T\Delta y^{\mathcal{D}}) = r_c \quad (25)$$
$(\lambda, \Delta y^{\mathcal{D}})$ denotes the inexact solution. We can rewrite (25) as:
$$[\mathbb{S}V + \mathbb{X}QV - \mathbb{P}\mathcal{G}^T] \cdot \begin{bmatrix} \lambda \\ \Delta y^{\mathcal{D}} \end{bmatrix} = r_c \quad (26)$$
System (26) has an $n \times n$ dimension and is guaranteed to be nonsingular. Using this system, the inexactness only manifests in $r_c$ and convergence is guaranteed. Therefore, we use this system for NT-QIPM since the primal and dual feasibility is preserved. The NT-QIPM is presented in Algorithm II.

---

**Algorithm II** NT-QIPM

1: choose $\epsilon > 0, \{\delta, \beta\} \in (0,1), \sigma = \left(1 - \frac{\beta}{\sqrt{n}}\right)$, and $k^{max}$
2: $k \leftarrow 0$
3: Initial feasible solution $((x^{\mathcal{P}})^0, (y^{\mathcal{D}})^0, (s^{\mathcal{D}})^0)$
4: **While** $((x^{\mathcal{P}})^0, (y^{\mathcal{D}})^0, (s^{\mathcal{D}})^0) < \varepsilon$ **and** $k < k^{max}$
5:     Calculate $\mu^k = M \times \frac{\left((x^{\mathcal{P}})^k\right)^T (s^{\mathcal{D}})^k}{n}$
6:     Convert (26) to the form of $\hat{\mathcal{A}}\hat{\mathcal{X}} = \hat{\mathcal{R}}$
7:     Send the matrix $\hat{\mathcal{A}}$ and vector $\hat{\mathcal{R}}$ as input of Algorithm I
8:     Calculate $(\lambda^k, (\Delta y^{\mathcal{D}})^k)$ using Algorithm I
9:     $(\Delta x^{\mathcal{P}})^k = V\lambda^k$ and $(\Delta s^{\mathcal{D}})^k = -\mathcal{G}^T(\Delta y^{\mathcal{D}})^k$
10:    $((x^{\mathcal{P}})^{k+1}, (y^{\mathcal{D}})^{k+1}, (s^{\mathcal{D}})^{k+1}) = ((x^{\mathcal{P}})^k, (y^{\mathcal{D}})^k, (s^{\mathcal{D}})^k) + ((\Delta x^{\mathcal{P}})^k, (\Delta y^{\mathcal{D}})^k, (\Delta s^{\mathcal{D}})^k)$
11:    $k \leftarrow k + 1$
12: **end**
13: **Return** $(x^{\mathcal{P}})^k, (y^{\mathcal{D}})^k, (s^{\mathcal{D}})^k$

---

In the context of the proposed quantum-assisted NT-QIPM, we aim to enhance the efficiency of Step 7 by leveraging quantum algorithms. Specifically, we use HHL to obtain the solution to the linear system of equations (26).

### B. Classically Augmented NT Quantum IPM

Although Algorithm II enhances the solution accuracy, it may still not provide satisfactory solutions under considerable quantum noise and error. To address this challenge, we propose a hybrid classically augmented NT-QIPM (CNT-QIPM) approach. We run NT-QIPM while monitoring its convergence behavior by calculating the following index, which is the average of the objective function ($\mathcal{M}_j$) improvement over the past $n$ iterations.

$$\mathcal{M}_j = \frac{\sum_{k=j-n}^{2 \times n-j} f_{k+n}}{n} \quad \forall j = 1,2,\ldots, k^{max} - (n-1) \quad (27)$$

$$Conv_j = \left|\frac{\mathcal{M}_{j+1} - \mathcal{M}_j}{\mathcal{M}_j}\right| \quad \forall j = 1,2,\ldots, k^{max} - (n-2) \quad (28)$$

$$I_j^{conv} = |Conv_{j+1} - Conv_j| \quad \forall j = 1,2,\ldots, k^{max} - (n-3) \quad (29)$$

where $f_k$ represents the value of the objective function in iteration $k$, and the index $Conv_j$ is indicative of the convergence speed. A slow convergence is underway if this index falls below a threshold ($\varepsilon_{conv}$) for two consecutive iterations. We stop NT-QIPM and use the result of the last iteration, which is expected to fall into a good central path toward the optimal solution, as a starting point for the classical IPM. In this case, the optimal solution is obtained after only a few classical IPM iterations. The proposed CNT-QIPM is presented in Algorithm III. This approach can obtain the optimal solution with a lower runtime than classical IPM. Most iterations belong to the NT-QIPM step rather than the IPM step. This reduces the CNT-QIPM runtime as the runtime of each iteration in classical IPM is approximately $O(n^3)$, whereas it is nearly $O(\log(n))$ in the case of NT-QIPM.

---

**Algorithm III** CNT-QIPM.

**Input:** Matrix $\mathcal{A}$ and column vector $\mathcal{R}$
**Output:** A solution vector $\mathcal{X}$
1: Run NT-QIPM
2: Compute $I_j^{conv}$ as the objective function improvement index
3: **If** $I_j^{conv} < \varepsilon_{conv}$
4:     Initialize IPM with NT-QIPM results of the last iteration
5:     Run IPM
6: **Elseif** Iteration number $< k^{max}$
7:     Continue NT-QIPM

**Return:** Exact solution

## V. CASE STUDY

### A. Simulation Settings

The performance of the proposed algorithms is analyzed on four test systems: a 3-bus system, a 5-bus system, the IEEE 118-bus system, and the IEEE 300-bus system. For the two smaller systems, the Qiskit module is utilized due to the manageable dimensions of the system of linear equations. Qiskit is an open-source quantum computing simulator that offers simulation capabilities with modeling noise and other error sources in quantum gates and circuits [31]. We use Qiskit default settings for noise and error modeling. To accurately assess the Qiskit output, a separate noise modeling is incorporated to account for hardware inaccuracies.

For the two larger systems, the primary focus is noise modeling to observe the error trend of NT-QIPM. The dimensionality of the system of linear equations poses difficulties for the current Qiskit modules to handle such a large matrix $\mathcal{A}$. We address this limitation by simulating HHL errors to examine the performance of NT-QIPM. Two approaches are considered to simulate errors. In approach 1, a constant error of 20% is applied to matrix $\mathcal{A}$. This allows for an assessment of the impact of encoding, phase estimation, and rotation stage errors in HHL on the accuracy of eigenvalues. In approach 2, we add a random uniform error within the range of $\pm 10\%$ to matrix $\mathcal{A}$. In both approaches, the IBM noise modeling module is incorporated into the solution at each iteration. This combination of errors serves to simulate a realistic noise and uncertainties present in real quantum hardware. We have run NT-QIPM for the 3-bus and 5-bus systems, as they are solvable by Qiskit. We have used error simulations of approaches 1 and 2 and repeated the simulations using Matpower within Matlab. The convergence patterns observed after implementing the error simulation approaches 1 and 2 (particularly the latter) follow the same trend as that obtained by modeling error with Qiskit. For the 118-bus and 300-bus system studies involving larger dimensions of linear systems of equations that are not solvable by the existing Qiskit version, we use the two error simulation approaches as representative of the Qiskit module and real quantum computer errors. These investigations contribute to understanding limitations and potential mitigations when applying NT-QIPM to solve DCOPF problems on a larger scale.

The classical IPM is executed on Matpower 7.1 toolbox in Matlab, and Python is used for HHL implementation.

### B. QIPM Performance Analysis

We solve DCOPF with QIPM by replacing (20) instead of (26) in Algorithm II. This change makes the algorithm QIPM instead of NT-QIPM. Table I shows the results. While we observe successful convergence in the 300-bus system, QIPM yields unsatisfactory results in the remaining cases. This can be attributed to the larger scale of the 300-bus system, which exhibits reduced sensitivity to errors and noise, facilitating its convergence. Consequently, considering these outcomes, the QIPM method lacks reliability for obtaining DCOPF solutions in the presence of quantum noise and error.

TABLE I
DCOPF USING QIPM

| System | Iteration | Convergence |
|---|---|---|
| 3-bus | 42 | Numerically failed |
| 5-bus | 21 | Numerically failed |
| 118-bus | 18 | Numerically failed |
| 300-bus | 12 | Converged |

### C. NT-QIPM Performance Analysis

We analyze the performance of NT-QIPM of Algorithm II for solving DCOPF. We conduct a comparative analysis of IPM and NT-QIPM.

*5-Bus System:* As shown in Fig. 5a, the exact IPM takes 16 iterations. NT-QIPM takes approximately 110 iterations to find a solution that resembles IPM's. Fig. 5b shows the difference between the objective function obtained by the two methods, which we call NT-QIPM error. The error remains below a threshold of 1.5% upon NT-QIPM convergence. This observation suggests that the difference between solutions obtained from the two methods is relatively small. To gain more insight, power generation and line flow values are presented in Fig. 6. The proximity between the solutions implies that, despite challenges in achieving explicit convergence within the defined tolerance level, the overall accuracy of NT-QIPM remains reasonably high.

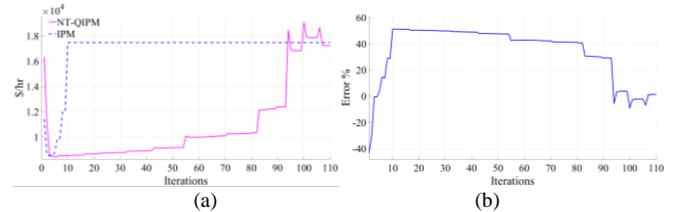

Fig. 5. a) 5-bus DCOPF objective value over IPM and NT-QIPM iterations, and b) error percentage.

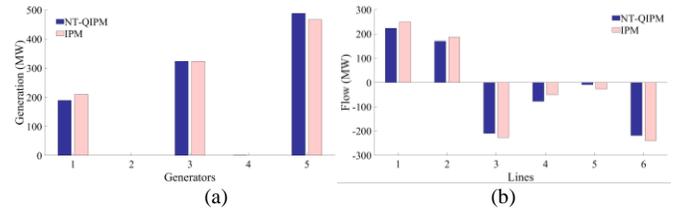

Fig. 6 a) Power generation and b) line flows of IPM and NT-QIPM.

To further study the NT-QIPM performance, we change the load within a range of 80% to 120% of its nominal value and evaluate the DCOPF objective function relative error as compared to IPM. Fig. 7 presents the results. The pattern for all cases is the same: error overshoots at the beginning and then moves towards error reduction and convergence. The overshoot is less for lower loads and increases as the load level grows. The results disclose the fine performance of NT-QIPM in various load levels and highlight its potential as a dependable approach for DCOPF even under uncertainties and fluctuations in power system loads.

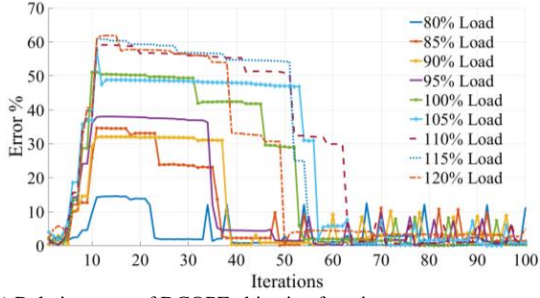

Fig. 7 a) Relative error of DCOPF objective function.

*3-Bus System:* Small systems are usually more sensitive to any changes in the system or numerical solution methodologies. For the 3-bus system, we expect an increased susceptibility to errors and deviations. This is because a relatively lower number of variables magnifies the impact of any inaccuracies or inexactness introduced during the analysis. Fig. 8a shows the objective function values obtained by NT-QIPM and the classical IPM after ten iterations. Fig. 8b represents the error percentage between the two methods, and Fig. 9 shows the generation and line flow values. NT-QIPM achieves an error of less than 5% after ten iterations. The convergence pattern shows a slow error reduction. For this small test system, the standard configuration of NT-QIPM may not initially yield a solution that closely approximates the optimal result. Nonetheless, by extending the maximum iteration number to a significantly higher value, the method exhibits a trend of decreasing error, eventually reaching an acceptable level of accuracy.

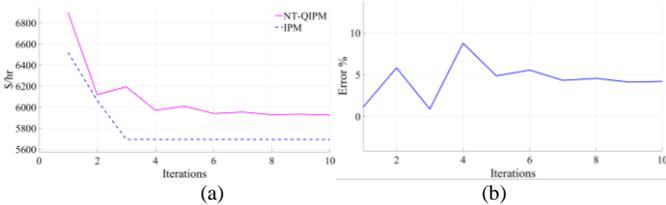

Fig. 8 a) DCOPF objective value over IPM and NT-QIPM iterations, and b) relative error percentage corresponding to NT-QIPM.

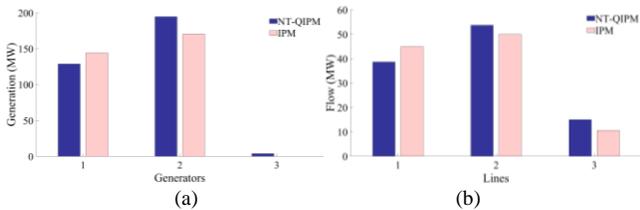

Fig. 9 a) Power generation and b) line flows of IPM and NT-QIPM.

*Two Larger Power Systems:* Due to Qiskit limitations, HHL cannot be run for the 118-bus and 300-bus systems. We simulate HHL error and quantum computing noise using error simulation approach 1. Fig. 10 demonstrates the DCOPF objective function value over the course of NT-QIPM iterations. NT-QIPM converges within a few iterations. Comparatively, when examining the 3-bus system, it becomes apparent that the sensitivity of NT-QIPM decreases as the system size increases. This phenomenon can be attributed to the characteristics of large-scale systems, where minor alterations do not significantly impact the convergence process.

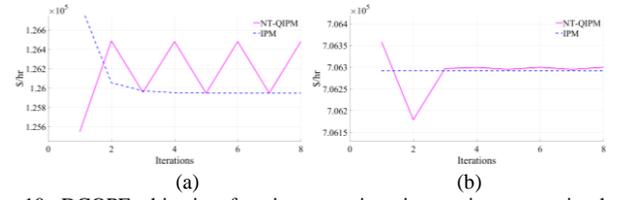

Fig. 10. DCOPF objective function over iterations using error simulation approach 1: a) 118-bus system and b) 300-bus system.

We have repeated the simulations using error modeling approach 2. Fig. 11 shows that NT-QIPM progressively approaches the optimal values as the number of iterations increases. Within a few iterations, NT-QIPM narrows the gap and achieves acceptable results close to the exact values obtained by the classical IPM.

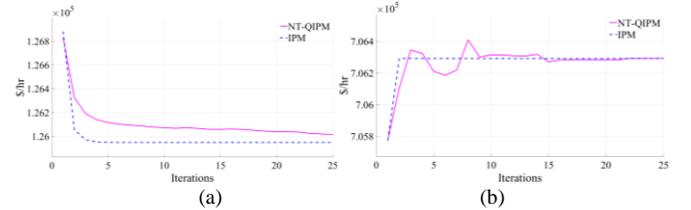

Fig. 11. DCOPF objective function using error modeling approach 2: a) 118-bus system and b) 300-bus system.

For further analysis, the value of objective function relative error under different load scenarios is examined on the 300-bus system. These comparisons allow for a comprehensive understanding of how the errors in the objective function vary depending on the specific load conditions and visualize the performance of NT-QIPM, contributing to an analysis of the method's effectiveness in practical applications. Fig. 12a illustrates NT-QIPM error under different load scenarios when applying a constant 20% quantum computing error (i.e., quantum error simulation approach 1). The number of iterations varies with load. In Fig. 12b, a similar behavior is observed when applying quantum error simulation approach 2, but with generally larger errors compared to Fig. 12a. This difference can be attributed to the use of random uniform errors in quantum error simulations. Despite the presence of errors, NT-QIPM exhibits the capability to approach the optimal values, although with slight variations depending on the specific load conditions.

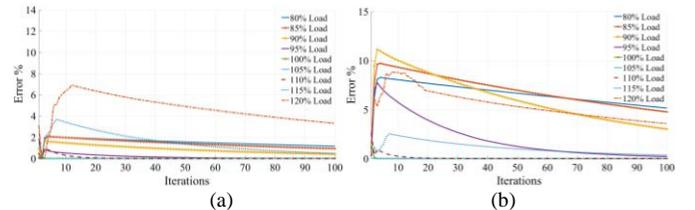

Fig. 12. 300-bus objective function relative error under various load levels: a) approach 1 and b) approach 2.

### D. CNT-QIPM Performance Analysis

We have observed that NT-QIPM demonstrated the ability to approximate the optimal value in various scenarios after a certain number of iterations. However, in some instances, such as those depicted in Fig. 8, Fig. 10, and Fig. 11b, NT-QIPM either became stuck in a repetitive pattern or exhibited a notably slow convergence toward the optimal solution. In addition, NT-QIPM usually needs a high iteration number, and in most cases, the objective function error remains high even under a relatively high iteration number, such as 50. Table II presents the relative percent errors of NT-QIPM solution under various load levels.

TABLE II
NT-QIPM ERROR AFTER 50 ITERATIONS

| System | Load level | | | | | | | | |
|---|---|---|---|---|---|---|---|---|---|
| | 0.8 | 0.85 | 0.9 | 0.95 | 1 | 1.05 | 1.1 | 1.15 | 1.2 |
| 3-bus | 3.6 | 0.8 | 1.4 | 0.3 | 0.2 | 0.1 | 0.1 | 0.7 | 0.0 |
| 5-bus | 6.6 | 7.5 | 6.5 | 1.6 | 0.1 | 0.0 | 0.1 | 0.8 | 5.8 |
| 118-bus | 1.9 | 1.4 | 2.4 | 10.7 | 2.9 | 11.8 | 12 | 29 | 15 |
| 300-bus | 1.2 | 22.4 | 3.4 | 3.3 | 4.1 | 10.6 | 16 | 13.8 | 3.3 |

To confront these errors, we utilize the proposed Algorithm III. We use the CNT-QIPM algorithm for 3-bus and 118-bus systems. The threshold $\varepsilon_{conv}$ is set to $10e-4$. Fig. 13 shows the result for the objectives function compared to IPM. Using NT-QIPM as a preliminary step, followed by passing the obtained values to the IPM, results in a drastically reduced number of iterations required to find the optimal solution. The solutions are identical to those obtained using IPM. To show the performance of this algorithm and its ability to expedite the convergence process of NT-QIPM, Fig. 14 depicts the convergence behavior of the proposed algorithm under various load levels for the 300-bus system, utilizing approach 2 for error simulation. In all scenarios, the error drops to 0% and the exact DCOPF solutions are obtained. Comparing Fig. 14 with Fig. 13b shows the desired convergence pattern of CNT-QIPM as compared to NT-QIPM.

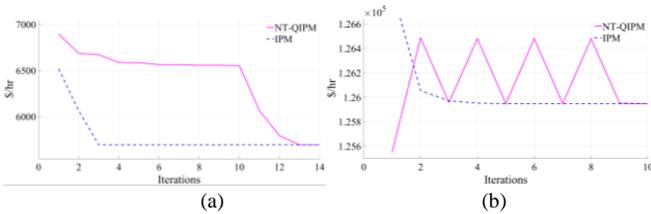

Fig. 13. CNT-QIPM-based DCOPF for a) the 3-bus system and b) the 118-bus system.

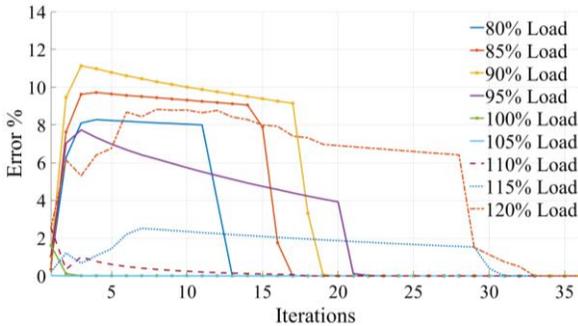

Fig. 14. CNT-QIPM-based DCOPF for the 300-bus system under various load levels.

*E. CNT-QIPM Runtime Analysis*

Compared to the classical IPM, the CNT-QIPM algorithm theoretically reduces the total running time to reach the optimal solution since IPM iterations consume a relatively longer duration than QIPM iterations performed by Algorithm III. This difference is due to the logarithmic time complexity of HHL, which is considerably faster than the polynomial time complexity of IPM. To assess the impact of this proposed approach, we have conducted experiments on four test systems with loads ranging from 80% to 120%. The number of QIPM followed by IPM iterations is presented in Table III. For instance, for the 300-bus system with the nominal load factor, CNT-QIPM takes 15 QIPM iterations flowed by two IPM iterations. The number of IPM iterations is considerably low for larger systems.

TABLE III NUMBER OF NT-QIPM ITERATIONS IN ALGORITHM III

| System | Load level | | | | | | | | |
|---|---|---|---|---|---|---|---|---|---|
| | 0.8 | 0.85 | 0.9 | 0.95 | 1 | 1.05 | 1.1 | 1.15 | 1.2 |
| 3-bus | 25 (6) | 20 (6) | 20 (6) | 20 (6) | 20 (5) | 20 (5) | 20 (6) | 20 (6) | 25 (7) |
| 5-bus | 35 (4) | 30 (3) | 20 (3) | 15 (3) | 15 (4) | 15 (4) | 15 (4) | 15 (5) | 15 (6) |
| 118-bus | 20 (3) | 20 (3) | 15 (2) | 15 (2) | 10 (2) | 10 (2) | 10 (2) | 10 (2) | 10 (3) |
| 300-bus | 20 (4) | 20 (3) | 15 (3) | 15 (3) | 10 (2) | 10 (2) | 10 (2) | 30 (4) | 35 (4) |

*Numbers in the parentheses show the number of iterations that classical IPM needs for convergence

Fig. 15, with the time axis on a logarithmic scale, shows the theoretical running time of solving the linear system of equations in IPM, NT-QIPM, and CNT-QIPM. While NT-QIPM is the fastest approach, its accuracy is less than CNT-QIPM due to quantum noise and HHL error. CNT-QIPM exhibits faster convergence compared to IPM. The time required for solving CNT-QIPM remains consistently minimal across all scenarios. Table IV presents the extent to which the theoretical running time of CNT-QIPM is faster than that of classical IPM. As the system size increases, the contrasting disparity between IPM and CNT-QIPM becomes evident, as IPM needs more time to converge, primarily due to its operation on a polynomial time scale. For the 300-bus system, for instance, CNT-QIPM is 449% faster than IPM. Using CNT-QIPM, we can obtain the same results as IPM in a shorter timeframe.

TABLE IV
SPEEDUP PERCENTAGE OF CNT-QIPM OVER IPM

| System | | | |
|---|---|---|---|
| 3-bus | 5-bus | 118-bus | 300-bus |
| 197% | 398% | 549% | 449% |

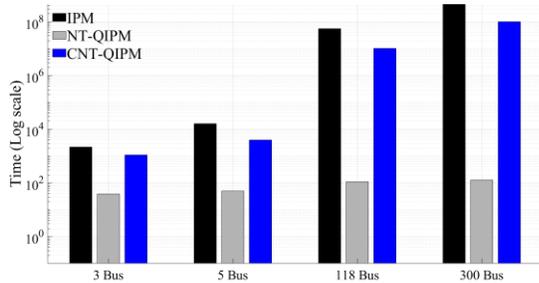

Fig. 15. Theoretical running time of IPM, NT-QIPM, and CNT-QIPM.

## V. Conclusion

This paper presents quantum computing-inspired algorithms for solving the DCOPF problem. We reformulate the DCOPF problem as a linearly constrained quadratic optimization and solve it by the primal-dual interior-point method. The HHL quantum algorithm, which offers a logarithmic time scale, is used to determine the Newton direction at each IPM iteration. However, it is important to acknowledge that HHL solves linear systems of equations with a certain level of inexactness. Therefore, we adopt a noise-tolerant QIPM to ensure feasibility and convergence. A hybrid quantum-classic strategy is proposed to accelerate the performance of the classical IPM while maintaining its accuracy. This approach, referred to as CNT-QIPM, overcomes the inherent inaccuracies associated with HHL. Remarkably, CNT-QIPM yields identical results to the traditional IPM but with significantly less computational time. This study considers the inherent noises associated with quantum computers and the errors related to HHL. These considerations ensure a comprehensive analysis of the proposed algorithms and provide a realistic understanding of their performance under various conditions.

Four power systems are used to examine the performance of proposed algorithms under various load levels and error conditions. Among these systems, two are characterized as small-size systems, whereas the remaining two are larger-size systems. This selection allowed for a comprehensive analysis of the algorithm's effectiveness and scalability across different system sizes. The simulations performed on Qiskit, along with error simulation modeling, show consistent convergence trends. NT-QIPM obtained a solution for DCOPF within an acceptable range of errors. Convergence behavior varies depending on the system characteristics and other conditions; however, after a certain number of iterations, the algorithm progressively approaches an $\varepsilon$-approximate of the exact solution. The iteration count required to achieve the desired solution is generally higher for small systems. This can be attributed to the heightened sensitivity of the algorithm when dealing with fewer variables or activating more inequality constraints. In the end, CNT-QIPM effectively solves the DCOPF problem accurately accuracy and theoretically demonstrates lower time complexity than the classical IPM approach. The proposed CNT-QIPM showcases its potential to provide efficient and accurate solutions for power system optimization.